\documentclass[12pt, letterpaper]{article}

\usepackage{amsmath}
\usepackage{dcolumn}
\usepackage[utf8]{inputenc}
\usepackage{graphicx}
\usepackage{amssymb}
\usepackage{setspace}
\usepackage{rotating}
\usepackage{algorithm}
\usepackage{algorithmicx}
\usepackage[noend]{algpseudocode}
\usepackage[export]{adjustbox}
\usepackage[biblabel]{cite}
\usepackage{graphicx}
\usepackage{enumitem}
\usepackage{float}

\usepackage{url}

\DeclareMathOperator*{\argmin}{arg\,min}

\title{\fontsize{14}{15}\bfseries
LSTM Hyper-Parameter Selection for Malware Detection: Interaction Effects and Hierarchical Selection Approach}

\author{
Mohit Sewak\\
\texttt{Microsoft R\&D, India}\\
\texttt{mohit.sewak@microsoft.com}
\and
Sanjay K. Sahay, Hemant Rathore\\
\texttt{BITS Pilani, Goa, India}\\
\texttt{\{ssahay, hemantr\}@goa.bits-pilani.ac.in}
}

\date{} 

\begin{document}
\maketitle

\begin{abstract}
Long-Short-Term-Memory (LSTM) networks have shown great promise in artificial intelligence (AI) based language modeling. Recently, LSTM networks have also become popular for designing an AI-based Intrusion Detection Systems (IDS). However, its applicability in IDS is studied largely in the default settings as used in language models. Whereas security applications offer distinct conditions and hence warrant careful consideration while applying such recurrent networks. Therefore, we conducted one of the most exhaustive work on LSTM hyper-parameters for IDS and experimented with ~150 LSTM configurations to determine its hyper-parameters’ relative-importance, interaction-effects, and optimal selection-approach for designing an IDS. We conducted multiple analysis of the results of these experiments and empirically controlled for the interaction effects of different hyper-parameters covariate level. We found that for security applications, especially for designing an IDS, neither similar relative importance as applicable to language models is valid, nor is the standard linear method for hyper-parameter selection ideal. We ascertained that interaction effect plays a crucial role in determining the relative importance of hyper-parameters. We also discovered that after controlling for the interaction-effect, the correct relative importance for LSTMs for an IDS are batch-size, followed-by dropout ratio and padding. The findings are significant because when LSTM were first used for language models, the focus had mostly been on increasing the number of layers to enhance performance.
\end{abstract}


\section{Introduction} \label{sec:introduction}
Long Short-Term Memory (LSTM) networks have become very popular for the design of artificial intelligence (AI) tools to make sequential decision in analytical domains requiring analysis of text, speech, time-series, etc \cite{lstm-language-modeling}. Recently, its applicability is also studied for design security applications like of Intrusion Detection Systems (IDS) for malware detection and prevention \cite{sahay2020evolution}. Generally, IDS that use LSTM networks have been designed mostly with the default options as used in (spoken) language models, or experimented with just a few configurations, mostly involving changes in number of LSTM layers and/or its units \cite{lstm-language-modeling}.

This is not strange as these network architectures were highly inspired from the work in language modeling, where the number of LSTM layers is invariably proven to be the most important factor for improving the performance of complex Natural Language Processing (NLP) models. However, besides the number of units and layers, the choice of other hyper-parameters is also very important for any application which is based on LSTM networks. Even for language model tasks like named-entity-recognition, part-of-speech tagging, etc. interestingly parameters besides the number of LSTM layers and its units were found more influential when a larger number of LSTM configurations were experimented \cite{lstm-50-configurations}. There exist many (neural) search-based algorithms \cite{lstm-parameter-search} that could identify the best hyper-parameters combination without searching (experimenting through) all the key search combinations. But a drawback of these methods is that many of these are mostly black-box models and do not give adequate insights of relative importance (main effect), and interaction effects of different levels of the hyper-parameter covariates. Also, such algorithm does not aid the development of a robust and intuitive hyper-selection approach, that could be basis of a simpler and efficient mechanism. 

Therefore, experimental evaluation of the LSTM network hyper-parameters is crucial for multiple reasons, but unfortunately not much of work in this field. In this field, currently the work in \cite{lstm-50-configurations} is the most exhaustive one for LSTM for language modeling tasks to-date. There are inherent differences between language modeling (word sequences analysis) and opcode sequence analysis. Also, an IDS that is based on opcode sequence analysis is highly complex, much more complex that average language models due to the sheer vocabulary size and the permutations of patterns in which it could be applied and its resulting outcome. Therefore, when LSTM networks are applied to an IDS with their default hyper-parameters as inspired by work in language models, it may not be optimal. 
Therefore, in this paper we conduct one of the most exhaustive series of experiments with different configurations of LSTM using the benchmark Malicia dataset \cite{Malicia}. We experimented with 149 configurations of different combination of LSTM hyper-parameters and their covariate levels. Further we apply techniques to empirically control the interaction-effect while analyzing the results from these experiments. In this study, we also investigated the interaction-effects from different levels of various hyper-parameter covariates and demonstrate how the results could have gone wrong without this control.

To the best of our knowledge, this is the most exhaustive work on LSTM hyper-parameter assessment, not just in the security domain, but for any domain, ever conducted. None of the earlier studies focused specifically on these critical aspects, and hence the results from them may not be relevant when more than one hyper-parameter’s level is altered as all the hyper-parameters work in tandem. 
Based on the findings of the interaction-effect, we also propose an approach of hierarchical selection of levels of different hyper-parameters. To the best of our knowledge such exhaustive comparison on LSTM’s hyper-parameter’s relative importance for malware detection is not available in any known literature. 

The main contributions of this paper are as follows:
\begin{itemize}
\item Analyzing the relative importance of each of the architectural parameters using benchmark Malicia dataset, viz. sequence/ padding length, embedding size, No. of LSTM layers, size of LSTM layer, dropout ratio, batch size and optimizer algorithm.
\item Proving the existence of interaction effect across different levels of various hyper-parameters covariates of an LSTM network and determining its impact.
\item Proposing an effective strategy to hierarchically determine the levels/ values of different hyper-parameters starting from the most influential hyper-parameter to avoid biases arising due to interaction-effects.
\item Hypothesizing and demonstrating that the most influential hyper-parameters for LSTM as applied to malware-detection are different from that in their conventional application as in language modeling.
\item Comparing the effectiveness of recurrent DL (LSTM) networks-based IDS with non-recurrent DL (MLP-DNN) networks based IDS that use similar base features (opcodes).
\end{itemize}

The rest of the paper is organized as follows. In section \ref{sec:realted-work} we cover related work on the applicability of LSTM/ RNN in IDS. Then, in section \ref{sec:relatedwork-lstm} we describe the LSTM and its units. The detail of the dataset and its pre-processing is covered in section \ref{sec:malware-data-used}. Section \ref{sec:modeling} describes the experimental setup and configurations. We then present and discuss our results in section \ref{sec:results} and finally conclude the paper in section \ref{sec:conclusion}.

\section{Related work} \label{sec:realted-work}
 
Recurrent Neural Networks (RNN), and its variants like the LSTM and Gated Recurrent Unit (GRU) are the deep learning (DL) algorithms used for analysis and classification of sequential data \cite{SewakTENCON, Sewak-DL-Introduction-DRL}. Most of the work on RNN and its variants are done in context of language analysis and modeling \cite{LSTM-Language}. The opcode sequence that represents the machine-language abstraction of any file also represents a sequence of data. Therefore, recently LSTM like networks are also used to analyse these sequences and determined if the sequence represent a malicious file or not. The main aspect to any form of analysis with deep-learning networks is the selection of hyper-parameters. For a complex recurrent network architecture like the LSTM there are many hyper-parameters that could be optimized. But unfortunately, in the case of LSTM most of the studies have covered the selection of number of LSTM layers \cite{LSTM-layer-selection} and its unit alone, and the coverage of other hyper-parameter selection is not sufficient. In this, the most exhaustive work done on hyper-parameter selection for the LSTM network is discussed in \cite{lstm-50-configurations}. However as far as the development of IDS with LSTM network is concerned, a specific context of IDS with LSTM hyper-parameter has been discussed in \cite{lstm-android-opcodes}. Earlier based on static-analysis, IDS have been proposed for malware detection \cite{rathore2020detection, Santos-2013}. These IDS involve methods ranging from classical machine learning  \cite{Santos-2013, rathore2020identification} to deep learning, and within deep learning, methods like Auto Encoders (AE), Multi Layer Perceptron (MLP) based deep-neural-networks \cite{SewakARES, DeepSign-2015} to recurrent-neural-networks (RNN) \cite{Pascanu-2015}. In the recent efforts, success has been achieved with different DL techniques like Deep Belief Networks \cite{DeepSign-2015} and Deep Neural Networks \cite{Saxe-2015}. RNN and its variants like Echo State Networks \cite{Pascanu-2015} LSTM \cite{opcode-lstm-2layer} and Gated Recurrent Unit \cite{microsoft-lstm} and combination of recurrent and convolutional neural networks \cite{Kolosnjaji-2016} for supervised learning and classification. RNN based Auto-Encoders 
have also been used for feature generation for downstream supervised learning mechanism.
Some of the work on RNN/ LSTM based models as discussed in \cite{microsoft-lstm, opcode-lstm-2layer} used a language representation of the malware as reflected in the extracted opcode sequences of the candidate file. In this, some of the salient features used for the malware detection by LSTM are the API call-sequences \cite{LSTM-api-call}, system call-sequences \cite{LSTM-system-call} file-headers \cite{LSTM-PE-header} and opcodes \cite{opcode-lstm-2layer}. Such implementations have been used for the platforms viz. Windows Portable Executable (PE) \cite{SewakTENCON, LSTM-PE-header, sewak2020doom, SewakDRLDO},  Android \cite{LSTM-system-call, rathore2020robust} and IoT devices \cite{LSTM-iot}, and the application of such techniques is targeted towards the detection of a specific malware \cite{LSTM-api-call} to identify the obfuscated malware \cite{LSTM-obfuscation}

In 2018, Jinpei Yan et. al. \cite{lstm-android-opcodes} used LSTM on opcode sequence-features and concluded that since opcode sequences are much longer than language representations, they may not work optimally for LSTM. In the works by \cite{opcode-lstm-2layer, LSTM-iot}, it has been found that LSTM with more number of layers work better for the malware detection. Also, \cite{lstm-pe-75000, lstm-android-1000-epochs} studied the application of LSTM for the malware-detection. But these studies are not on a benchmark dataset and hence could not be used for any comparison. Moreover the focus of their studies is not to explore the significance of hyper-parameter of LSTM as applied to malware-detection and \cite{lstm-android-1000-epochs} used very specific platform-based featured (Android permissions), which are not comparable to opcodes either in complexity or sequence-length. Hence, we understand that there exists the following gaps in the prior art that this paper intends to fill:
\begin{itemize}
    \item No exhaustive work exists that identify the relative importance of different configurable architectural and hyper-parametric settings of LSTM for an IDS or even in security domain in general for any security related problem. 
    \item Even outside of security domain, there exists no work to study the interaction effect of different hyper-parameters of an LSTM network.
    \item Also, as per our knowledge prior work exist that compares MLP-DNN based IDS directly with LSTM based IDS on a given benchmark dataset.
\end{itemize}

\section{Long Short Term Memory and its Units} \label{sec:relatedwork-lstm}
 Long Short Term Memory is a type of RNN, and recently it become very popular to analyse the sequential data (e.g. text, speech, time-series, etc.). This is because, LSTM works on the sequence of input features to predict the next likely outcome. Typically an LSTM unit consists of a cell that has an \textit{input}, \textit{output} and a \textit{forget} gate. These gates determine which information to learn, to remember or to forget across different steps in the sequence. Owing to this regulatory action of these gates, the LSTM cell is capable of remembering information over arbitrary time intervals, and therefore it is able to solve the basic problem that RNN suffers, i.e. \textit{Exploding} and \textit{Vanishing Gradient} problem. Because of these drawbacks the basic RNN is not able to effectively discover complex patterns in long sequences of data. The problem is similar to the classical MLP based DNN that had suffered before the advances were made in the activation function (Rectified Linear Unit). 
 \par

Since LSTM unit has two type of memory states, namely the long-term memory and the short-term memory, hence, it is able to strike a better balance between information that influence the outcome in different subsequent lengths of forthcoming sequences and fades off at different rates. In LSTM the \textit{hidden-state vector} $h_t \in \mathbb{R}^h$ at the sequence/time step `t' is given as:
\begin{equation} \label{eqn:lstm-hidden}
     h_t = o_t \circ \sigma_h(c_t)
\end{equation}
 
 \noindent where, the symbol `$\circ$' represents the \textit{Hadamard-Product} \cite{lstm-hadamard-product} which is a form of an element-wise product.  $\sigma_h$ is the hyperbolic-tangent function \cite{lstm-peephole-activation}, and $o_t \in \mathbb{R}^h$ is the output gate's \textit{state-vector} for an input vector $x_t \in \mathbb{R}^d$ at step `t' in the forward pass and is given as:
  \begin{equation}
     o_t = \sigma_g(W_ox_t + U_oh_{t-1} + b_o)
     \label{eqn:lstm-output}
 \end{equation}
 where, $W\in\mathbb{R}^{h\times d}$, and $U\in \mathbb {R} ^{h\times h}$ represents the weight-matrices
 and, $c_t$ is the \textit{cell's state-vector} and can be given as:
  \begin{equation}
     c_t = f_t \circ c_{t-1} + i_t \circ \sigma_c(W_cx_t + U_ch_{t-1} + b_c)
     \label{eqn:lstm-cell}
 \end{equation}
Here, the cell state-vector uses the \textit{hyperbolic-tangent} activation-function \cite{sewak-dl-overview} represented by  $\sigma_c$. The  $i_t \; \text{ and } \; \\f_t \in \mathbb{R}^h$ are the \textit{input} and  \textit{forget}  gate's \textit{state-vector} at step `t' respectively and can be given as:
  \begin{equation} 
     i_t = \sigma_g(W_ix_t + U_fh_{t-1} + b_i)
     \label{eqn:lstm-input}
 \end{equation}
\begin{equation}
    f_t = \sigma_g(W_fx_t + U_fh_{t-1} + b_f)
    \label{eqn:lstm-forget}
 \end{equation}
In the above, $b_o$, $b_c$, $b_i$ and $b_f$ are the bias(es) of $o_t$, $c_t$, $i_t$ and $f_t$ respectively, and the forget, input and output gates uses the \textit{Sigmoid Activation} function $\sigma_g$ \cite{sewak-dl-overview}.

\section{Dataset and its pre-processing} \label{sec:malware-data-used} \label{sec:data-preprocessing}
\begin{figure*}[htb]
    \centering
    \includegraphics[width=\textwidth, height=2.8in]{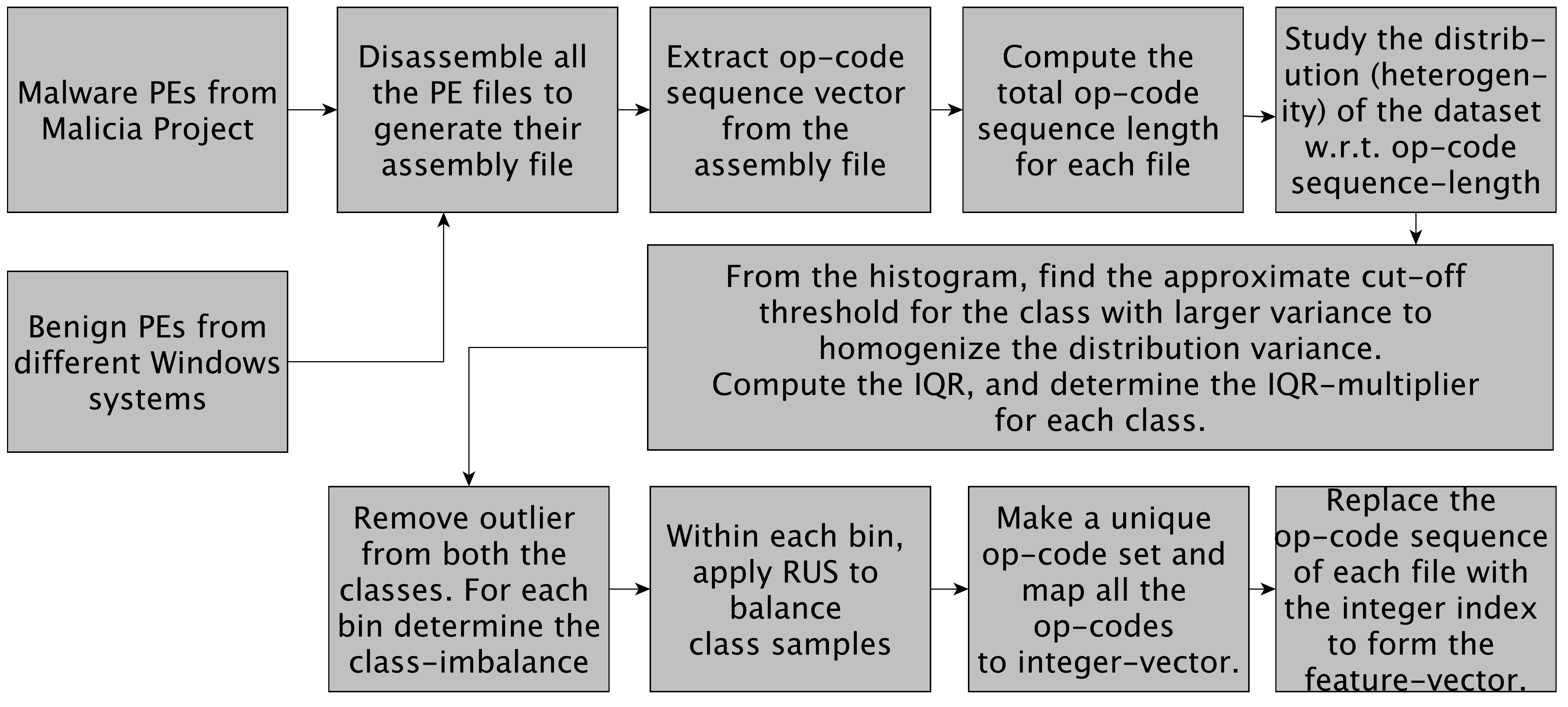}
    \caption{A schematic of the dataset pre-processing.}
    \label{fig:data-processing-flow}
\end{figure*}

Figure \ref{fig:data-processing-flow} depicts the flow of the dataset pre-processing for the assessment of the importance of different hyper-parameters of LSTM and to investigate how relevant will be if instead of applying frequency opcode vector (using non-recurrent DL), opcode sequence-vector (using recurrent DL) is used for an IDS. For the experimental analysis we selected a standard and prominently used Malicia dataset \cite{Malicia} which is extensively used for the development of malware-detection models for Windows PE \cite{SewakARES}.
For the analysis, as we will also need benign Windows PE files. Hence, we collected $2,819$ benign PE from different windows systems, including the \textit{Cygwin} utility. To confirm that the collected benign PE were indeed benign, we cross-verified each PE using the \textit{virustotal.com} file scanner services. 
From the collected $11,368$ malicious PE from Malicia dataset and  $2,819$ benign PE, we were able to successfully disassemble $9,336$ malicious and $2,607$ benign files respectively (rest were discarded) using Unix’s \textit{objdump} utility and then extracted the opcode sequence vector from the disassembled files.

\begin{figure*}[htb]
    \includegraphics[width=\textwidth]{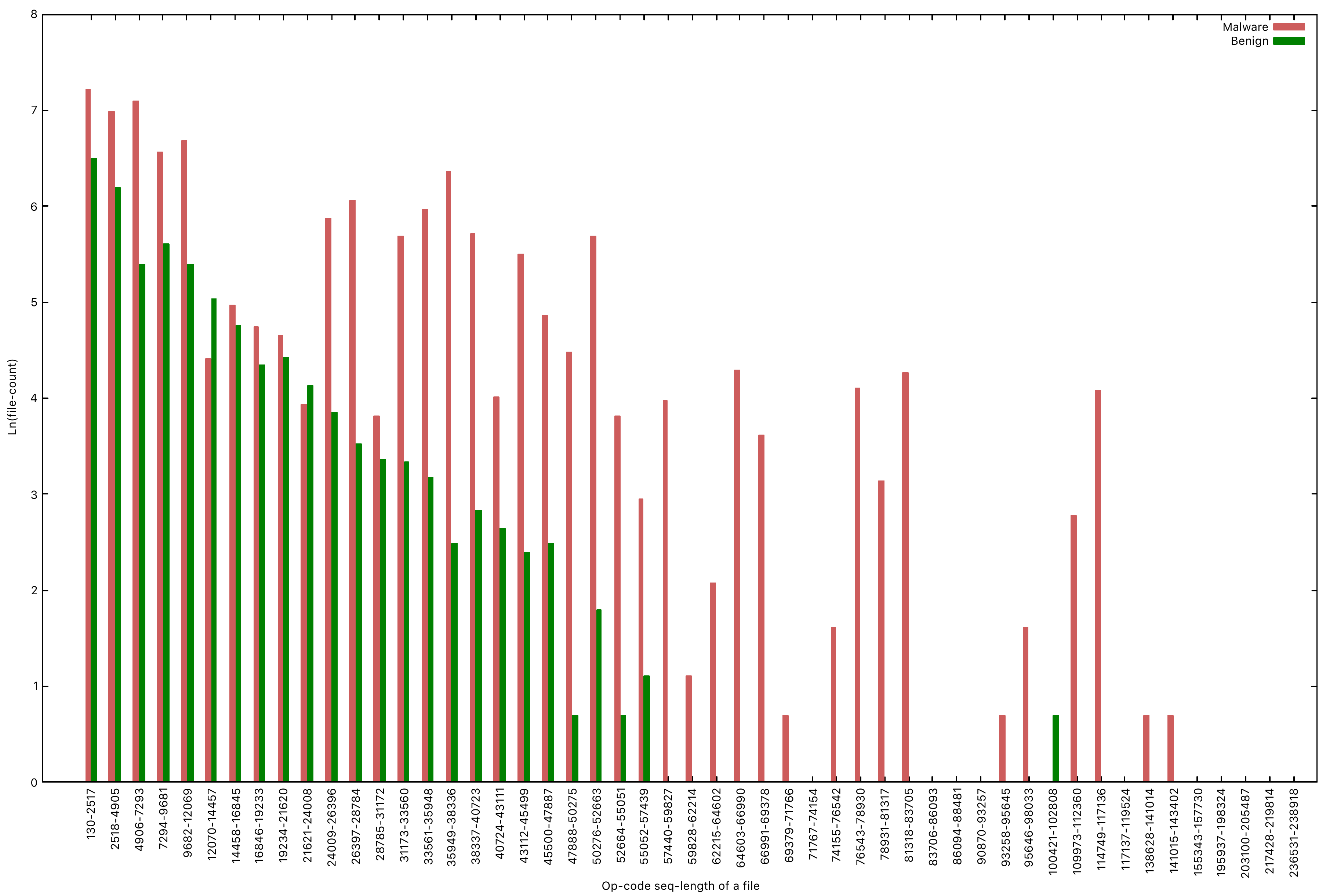}
    \caption{Distribution of malicious and benign files w.r.t opcode sequence-length in each file}
    \label{fig:histogram-opcode-count}
\end{figure*}

Since to train the LSTM model, the sequence length of the entire batch has to be homogeneous \cite{lstm-padding-length}, therefore to make the sequence-length (i.e., total-opcode count in each file) homogeneous for the analysis, first we computed the total-opcode-count in each file, and find that a maximum of $2,38,918$ and $1,02,391$ opcodes sequence-length are in a given malicious and benign file respectively. With so much heterogeneity in the opcode sequence-length between the two classes, it is challenging to determine the right sequence length to train the LSTM that optimally fits samples of both the classes. Additionally, we have to also reduce some malicious files to achieve balance between the sample size between the two classes.  Therefore, to homogeneous and balance the sample size between the two classes, we studied (Fig. 2) the distribution of opcode sequence-lengths in malicious and benign file by dividing the entire opcode range in $100$ bins, each of equal sized comprising a range of $2,387$ opcode sequence-lengths. Since the variation in the number of files in each bin is very large, therefore instead of taking absolute frequency (number of files in a bin), we have taken log of the occurrence of file in each bins. From the histogram (figure \ref{fig:histogram-opcode-count}), we observe a long-tails in the distribution of malicious files sequence-length and a significant imbalance in the number of malicious and benign files in each bin. Also, we find that after $24^{th}$ bin the number of benign PE are negligible. Hence, for the homogeneity and to strike a balance between the two classes we use inter-quartile range (IQR) for anomaly detection given as

\begin{equation}
     \text{OpCodeThreshold}_{M} = \mathbb{Q}2_M + 1.5\times(\mathbb{Q}3_M-\mathbb{Q}1_M)
    \label{eqn:opcode-anomaly-detection-malicious}
\end{equation}
\begin{equation}
     \text{OpCodeThreshold}_{B} = \mathbb{Q}2_B + 5.0\times(\mathbb{Q}3_B-\mathbb{Q}1_B)
    \label{eqn:opcode-anomaly-detection-benign}
\end{equation}
\noindent separately for the malware and benign dataset by removing the files on the upper side of the median of the data. Here, $\mathbb{Q}1$, $\mathbb{Q}2$, and $\mathbb{Q}3$ are the $1^{st}$, $2^{nd}$, and $3^{rd}$ quartile of the `total-opcode-count' distribution respectively, and $(\mathbb{Q}3-\mathbb{Q}1)$ is the IQR \cite{box-plot}.
In the above the IQR \textit{multiplier} for malicious is kept as $1.5$, but the IQR \textit{multiplier} for benign has been made more conservative and kept as $5.0$. This has been experimentally determined to achieve similar maximum sequence-lengths across the two classes. We use $\mathbb{Q}2 + \text{multiplier} \times$ IQR instead of the standard $\mathbb{Q}3 \; \pm \; \text{multiplier} \times$ IQR because histogram is right skewed and we want to remove the files on the upper side of the median only.

After correcting the skewness as above, we were left with $8,905$ malicious PE and $2,599$ benign PE, and  $56,909$ and $58,077$ maximum opcode sequence-length in malicious and benign data set respectively,  i.e. now the opcode sequence-length between the two classes are almost homogeneous. However, we are still left to attain homogeneity in the number of malicious and benign files in each bin. Therefore, in each selected bin we do random under-sampling (RUS) of the majority classes, and left with a total of $2,508$ malicious and benign files. RUS is done
because we do not want a negative impact on the distribution homogeneity and instead would prefer to impact it positively, we do a class balancing across these two classes individually in each of the $24$ bins. The distribution of the remaining file across the different bins are shown in table \ref{tbl:mal-ben-pe-in-bin}.

Now, after pre-processing the dataset, we were left with $905$ unique opcodes across all malicious and benign PE from the original $1,612$ unique opcodes \cite{SewakARES}. Therefore, for the experimental analysis, we made a opcode-set that contains $905$ unique opcodes, and then map all the $905$ unique opcodes to a contiguous sparse integer vector. Then we transform the opcode-sequence feature-vector obtained from the reduced dataset of each file by replacing the opcodes in the sequence vector with their respective integer-index from the opcode-integer mapping set.
\begin{table}
    \begin{center}
    \caption{No. of malicious/ benign PE in each bin.(M:Malware, B:Benign)}    
     \begin{tabular}{||c | c || c | c || c | c ||} 
     \hline
     Bin\#&M/B&Bin\#&M/B&Bin\#&M/B\\[0.5ex] 
     \hline\hline
    1&663&9&83&17&17\\
    2&486&10&51&18&14\\
    3&219&11&47&19&11\\
    4&271&12&34&20&12\\
    5&219&13&29&21&2\\
    6&82&14&28&22&6\\
    7&116&15&24&23&2\\
    8&77&16&12&24&3\\
    \hline\hline
    \end{tabular}
    \end{center}
    \caption{No. of malicious/ benign PE in each bin.}
    \label{tbl:mal-ben-pe-in-bin}
\end{table}
\raggedbottom

\section{Experimental setup and configurations} \label{sec:modeling}

\begin{figure*}[htb]
    \centering
    \includegraphics[width=\textwidth]{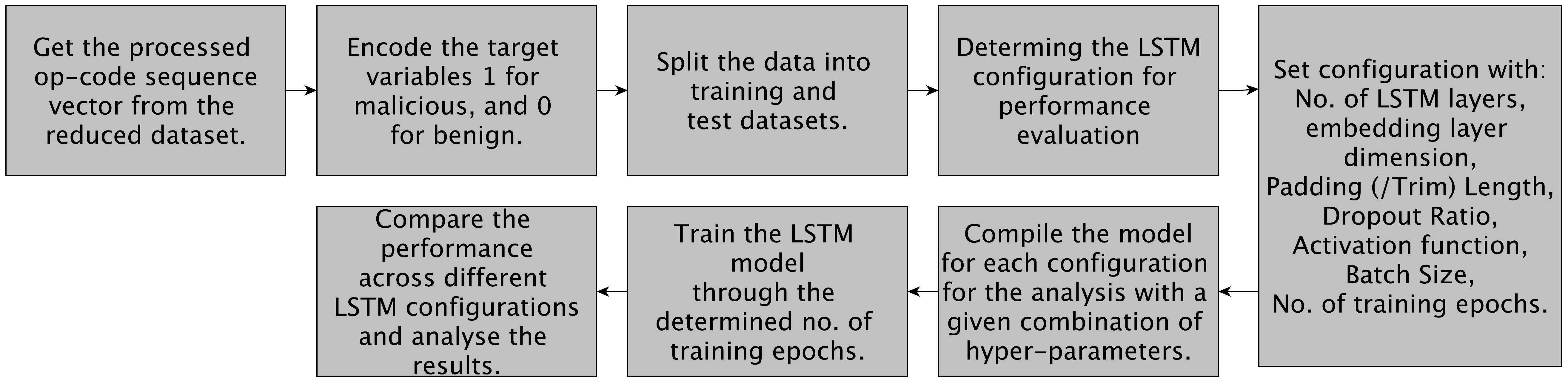}
    \caption{Process flow of the experimental analysis.}
    \label{fig:lstm-modeling-flow}
\end{figure*}

Figure \ref{fig:lstm-modeling-flow} shows a brief approach of the our investigation. For the experimental analysis first the pre-processed data set has been assigned with Boolean flag [0, 1] for each opcode sequence vector (0 for benign and 1 for malicious file), then we split the dataset into training ($70\%$) and test set ($30\%$) as a common resource for all the experiments to nullify the bias. Next, for each experiment we determine the configuration to be tested. This configuration comprises of a combination of hyper-parameter values/ levels from the set of hyper-parameters for which the importance has to be investigated. The hyper-parameters could be divided into two parts, i.e. the model-configuration parameters and training-configuration parameters. The model-configuration parameters determine the design of the model, and needs to applied before the model is compiled for training. The training-hyper-parameters on the other hand determine the training conditions with respect to the input data size and iterations. Once the model is defined as per the selected model-configuration-parameters (individual hyper-parameter levels except training hyper-parameters), the model is compiled so that the computation could be optimized and frozen for the training. Then input is fed to the model as per the training-hyper-parameters. At the end of each epoch and after determining the number of training epochs, the model is evaluated on the test dataset, and losses on this dataset is computed as validation loss (binary cross-entropy loss). Since accuracy and false positive rate are a function of the probability based class cut-offs for a given model, hence are not directly comparable across different models. Therefore we compare the models on the basis of their loss on validation dataset. The loss is further standardized to avoid any bias. Additionally we compute range and standardized-variance for comparing standardized-loss across hyper-parameters and range for comparing different levels of a single hyper-parameter as measures of dispersion. The LSTM network has been investigated with different number of LSTM layers, $N_{layers} \in \{1, 2, 3, 4\}$. For all LSTM layers we altered the number of LSTM units, $N_{units} \in  \{32, 64, 128\}$, and also the drop-out ratio, $\mathbb{R}_{dropout} \in  \{0.0, 0.1, 0.2, 0.3\}$. The drop-out ratio indicates the proportion of nodes that are rendered inactive during a forward-pass during training, and provides regularization like effect to avoid over-fitting in DL. A drop-out ratio of $0.0$ indicates no drop-out. Irrespective of the drop-out ratio, all nodes remain active during the validation forward-pass. 

\begin{figure}[htb]
    \centering
    \includegraphics[width=0.65\textwidth, rotate=90]{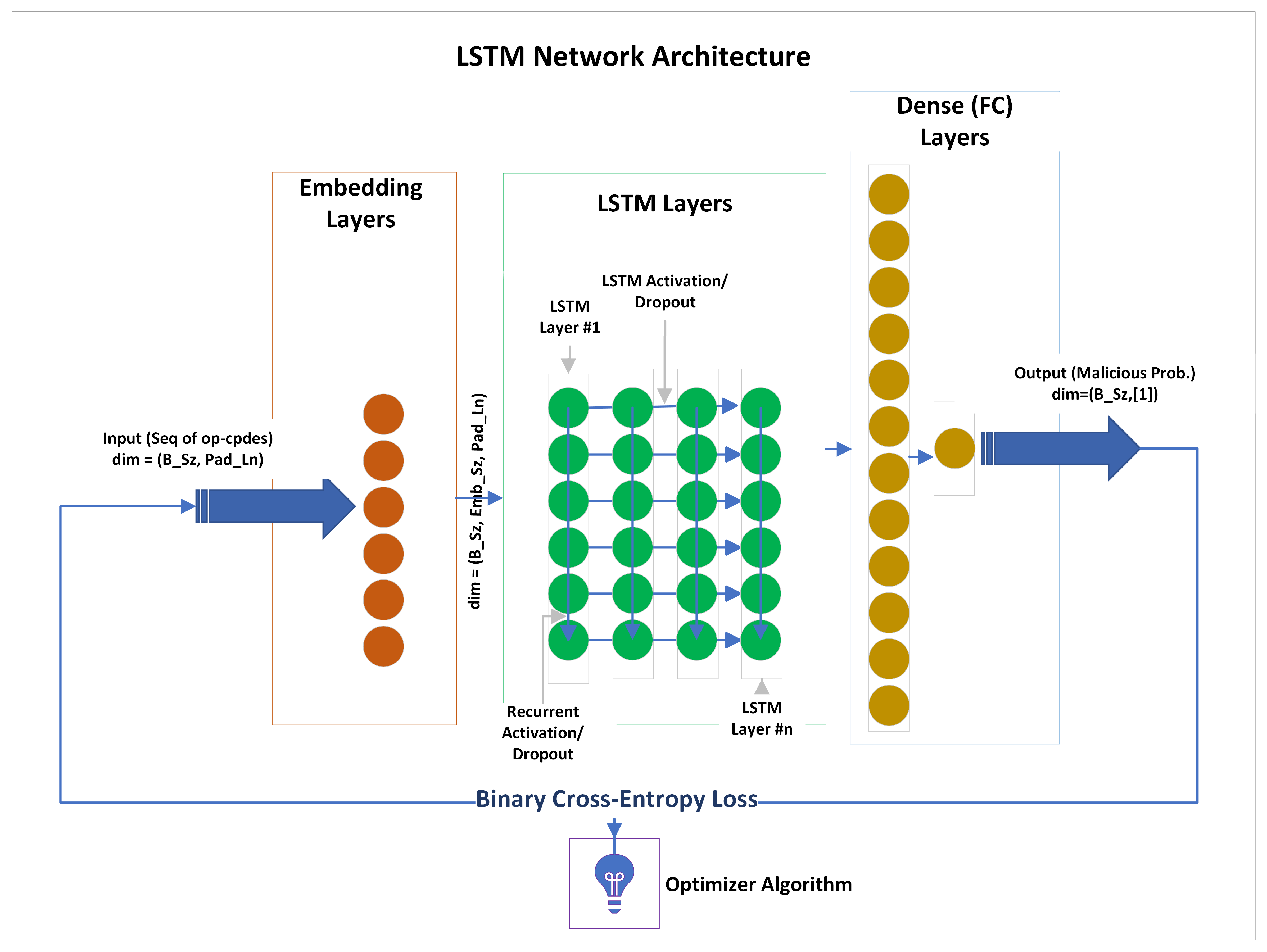}
    \caption{Architecture of the LSTM network for different configurations.}
    \label{fig:lstm-network-architecture}
\end{figure}  

As shown in the architecture of the LSTM network (figure \ref{fig:lstm-network-architecture}), there are two types of dropout parameters in LSTM, first is the dropout for the connection that goes out of an LSTM layer to another layer that succeeds it, second if the dropout of the recurrent connections that goes to the next sequence in the same layer. We altered both the dropout ratio simultaneously setting them to $\mathbb{R}_{dropout}$. The activation function for output to next layer, $Activation_{LSTM}$, and that for the recurrent connection, $Activation_{recurrent}$, has been kept as \textit{tanh}, and \textit{sigmoid} respectively in all the experiments. The first LSTM layer is preceded by an embedding layer to reduce the dimensionality to eliminate the sparsity in the sequence input. The mapped opcode sequence goes into the embedding layer and gets transformed into dense lower-dimensional representation, which then goes inside the LSTM. We vary the number of units, $N_{embedding} \in \{16, 32, 64, 129\}$, in the embedding layer, to alter the dimensionality of the dense representation output from this embedding layer.
Since LSTM require a fixed input-sequence/ padding length \cite{lstm-padding-length}, therefore we pad/trim the opcode input sequence before submitting them to the network after mapping it to different percentiles of the opcode sequence-length series of all the files instead
of using arbitrary padding length. We vary the $L_{padding}$ to $25^{th}, 50^{th}, 75^{th}, \text{and } 100^{th}$ percentile. Henceforth, these are prefixed with character $\mathcal{P}$, e.g., $\mathcal{P}_{25}$ represents $L_{padding}$ equal to the $25^{th}$ percentile of the sequence-length series.

As LSTM layers are succeeded by two dense/ fully-connected (FC) layers. Therefore, first FC layer has $256$ units and \textit{ReLU} activation, and the second has only one unit which is used to generate probability of the input sequence representing a malicious PE. For binary classification, this layer had a sigmoid activation and \textit{binary-cross-entropy} loss. To minimize the network loss we used either \textit{Adam} or \textit{RMSprop} optimizers, which are the some of the most popular optimizers used in DL \cite{sewak-dl-overview}. Now the opcode sequences from multiple PE were submitted together in batches. Each opcode sequence from a PE represents one record in the batch, and for a given batch size, $B_{size} \in \{64, 128, 256, 512, 1024\}$, the iterations per epoch is determined as $\lceil \frac{N_{PE}}{B_{size}} \rceil$. Since the objective of the paper is not to develop a fully trained IDS, but to evaluate the configurations, therefore the training was repeated only for $N_{epochs}=5$ epochs for all configuration, and the configurations were not trained till convergence. We conducted the experiments on \textit{Nvidia} $K80$ GPGPU, which has a memory limitation of $24$ GB, therefore configurations with a combination of hyper-parameters simultaneously involving higher \{$B_{size}$, $L_{padding}$, $N_{embedding}$, $ N_{units}$, $N_{layers}$\} could not tested. The training has been conducted using single-precision to include configurations with larger memory footprint to avoid bias.

\section{Results and Discussion} \label{sec:results}

\begin{figure}[htb]
    \centering
    \includegraphics[scale=0.5, rotate=-90]{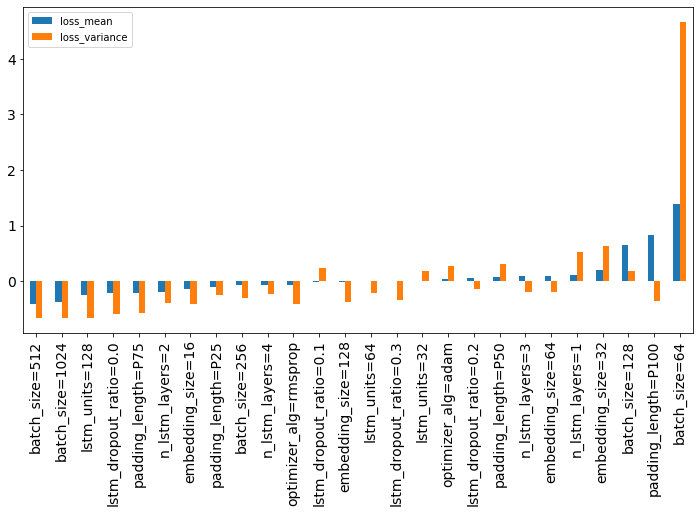}
    \caption{Mean-standardized-loss and variance of different levels of all the hyper-parameters}
    \label{fig:mean-loss-variance-all-parameters}
\end{figure}

Of all the permissible combinations of the different levels of all the hyper-parameters, $149$ configurations were found to be either suitable/ feasible for investigation. Therefore, $149$ configurations were experimented/ trained and the obtained results of these configurations ($N_c$) are shown in figure \ref{fig:mean-loss-variance-all-parameters}. For each hyper-parameter ($H$) we compute the mean of standardized (binary cross-entropy) loss, $L^s$ across all the configurations ($C$) such that for particular level ($H_L=L$) of the given hyper-parameter shall be observed ($L_{H_L=L}^s$), and can be represented by the equations
\begin{eqnarray}
        L^s & = & \frac{L_{C}-\Bar{L_{C}}}{\sigma_{L_C}} \\
        L_{H_L=L}^s & = & \mathbb{E}_{(N_C|H_L=L)} (L^s | H_L=L )
    \label{eq:mean-standardized-loss}
\end{eqnarray}

For a hyper-parameter to be influential for a particular problem, for at least some of its level, it should be capable to minimize the mean-standardized-loss. This is the primary criteria for ascertaining any hyper-parameter to be important, changing the levels of that parameter should significantly alter the $L^s$. If the $L^s$ is low for a given level of a hyper-parameter, altering it does not alter the $L^s$, then the low $L^s$ could also be due to the interaction effect of different hyper-parameter.

\begin{figure}[htb]
    \centering
    \includegraphics[width=\linewidth, height=2.5in]{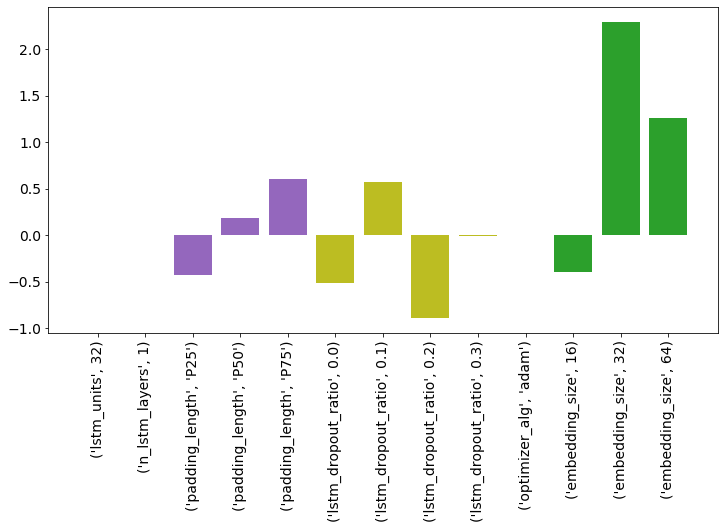}
    \caption{Mean Standardized Loss $| B_{size}=512$} 
    \label{fig:loss-given-bsize=512}
\end{figure}

From the analysis (figure \ref{fig:mean-loss-variance-all-parameters}) we observe that at $B_{size}=512$, the mean loss was least, followed by $B_{size}=1024$. But since $L_{B_{size}=1024}^s$ is slightly higher than $L_{B_{size}=1024}^s$, we can infer that $N_{epochs}=5$ are not sufficient to significantly reduce the loss in the reduced $N_{iteration}$ resulting from higher $B_{size}$. We also find that $B_{size}$ demonstrate the highest dispersion. With this we infer that for long and complex-sequence analysis as in the case of opcode sequences, batch-size is the most important hyper-parameter. 

\begin{figure}[htb]
    \centering
    \includegraphics[width=\linewidth, height=2.5in]{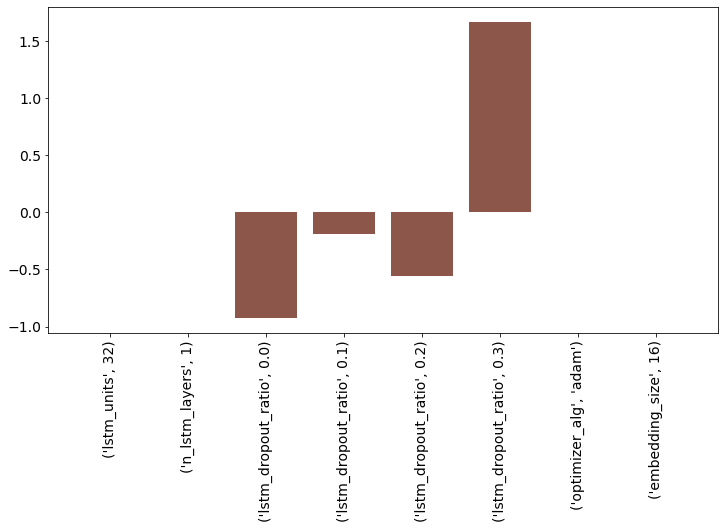}
    \caption{Mean Standardized Loss $| B_{size}=512, L_{padding}=\mathcal{P}_{25}$} 
    \label{fig:loss-given-bsize=512,pad=p25}
\end{figure}

To determine the relative importance of the remaining parameters is not very straight forward. This is due to interaction effect from hyper-parameters of higher importance. Therefore, to analyse the relative importance of the next set variables, we control the effects of $B_{size}$, and conduct the analysis in configurations $C | B_{size}=512$ and the obtained result is shown in figure \ref{fig:loss-given-bsize=512}. We find that $\argmin({L^S})$ is attained for the hyper-parameter $\mathbb{R}_{dropout}$, but highest range (dispersion) is demonstrated across the different levels of hyper-parameter $L_{padding}$. Since $\argmin (L^S)$ is the primary criterion, we can say that $\mathbb{R}_{dropout}$ is the second most important hyper-parameter. This could also be ascertained by controlling the third parameter (figure \ref{fig:loss-given-bsize=512,pad=p25}) and then observing the effects of $\mathbb{R}_{dropout}$ and $L_{padding}$. Analysis shows that $L^S$ conditioned-on/ controlled-for both $B_{size}=512, L_{padding}=P_{25}$, and confirms the earlier inference that $\mathbb{R}_{dropout}$ is the second most important hyper-parameter after $B_{size}$.

\begin{figure}
    \centering
    \includegraphics[width=\linewidth, height=2.5in]{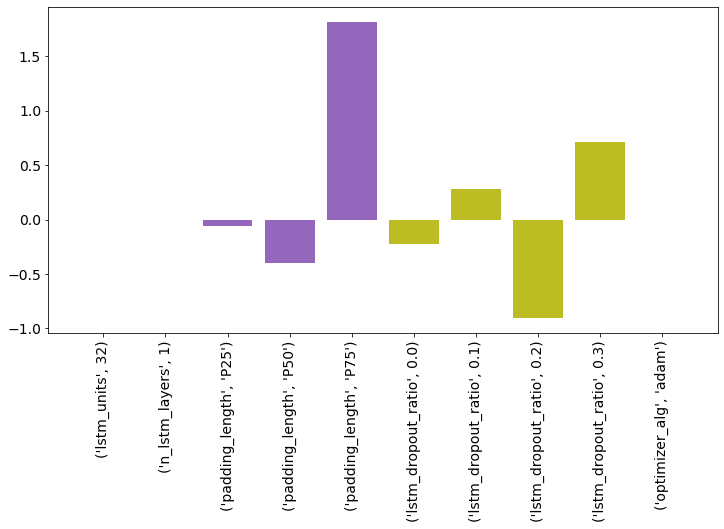}
    \caption{Mean Standardized Loss $| B_{size}=512, N_{embedding}=16$} 
    \label{fig:loss-given-bsize=512,emb=16}
\end{figure}

Next, we determine the level at which the parameter $\mathbb{R}_{dropout}$ generates the lowest mean-standardized-loss, and the results obtained is shown in figure \ref{fig:loss-given-bsize=512,emb=16}. We find that $L^S$ conditioned-on $B_{size}=512, N_{embedding}=16$, which confirms the fact that $\argmin L^S$ is at $\mathbb{R}_{dropout}=0.2$. Interestingly it also shows that $\argmin (L^S | H=L_{padding})$ is not at $L_{padding}=P_{25}$ as seen in figure \ref{fig:loss-given-bsize=512}.

\begin{figure}
    \centering
    \includegraphics[width=\linewidth, height=2.5in]{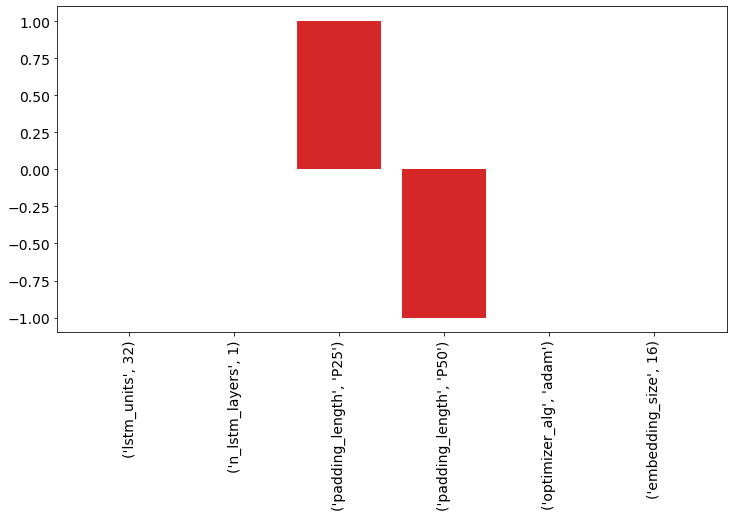}
    \caption{Mean Standardized Loss $| B_{size}=512, \mathbb{R}_{dropout}=0.2$} 
    \label{fig:loss-given-bsize=512,do=0.2}
\end{figure}

Now, for the third most important hyper-parameter $L_{padding}$ we determine the level at which it generates the lowest mean-standardized-loss, and the results obtained is shown Figure \ref{fig:loss-given-bsize=512,do=0.2}, which shows that $L^S$ conditioned-on $B_{size}=512, \mathbb{R}_{dropout}=0.2$, and also confirms the fact that $\argmin (L^S | H=L_{padding})$ is at $L_{padding}=P_{50}$.

The complete results of a sample configuration with hyper-parameters and their respective levels/ values as discussed in table \ref{tbl:configuration-with-selected-hyper-parameter} is given in the table \ref{tbl:performance-metrics-of-configuration-with-selected-hyper-parameter}. Finally, to understand the relevance of LSTM for IDS, a result obtained by Sewak et. al. is given in the table \ref{tbl:ae-dnn-results-120epochs} \cite{SewakARES}. They had achieved very high accuracy by using Auto-Encoders and DNN in on opcode frequency-vector on the same dataset. Though these configurations are trained for $N_{epochs}=120 \; (B_{size}=64)$, and the LSTM ones are trained for just $N_{epochs}=5$, but the huge gap in accuracy indicates the fact that the opcode sequence-vector obtained from the dataset is too complex to compete with the simpler frequency-vector. Also, the patterns across the combinations of opcodes that may lead to the detection of malicious intent may have patterns which could not be easily learn from the  dataset.



\begin{table}[htb]
    \centering
    \caption{LSTM configuration with selected hyper-parameters and their respective levels}
    \begin{tabular}{||c|c || c|c ||}
    \hline
         Parameter & Value & Parameter & Value\\
         \hline
$N_{embedding}$ & 16 & $N_{layers}$ & 1 \\
$L_{padding}$ & $P_{25}$ & $N_{units}$ & 32\\
optimizer alg & adam & $\mathbb{R}_{dropout}^{LSTM}$ & 0.2\\
$B_{size}$ & 512 & $\mathbb{R}_{dropout}^{Recurrent}$ & 0.2\\
\hline
    \end{tabular}
    \label{tbl:configuration-with-selected-hyper-parameter}
\end{table}
\raggedbottom

\begin{table}[htb]
    \centering
    \caption{Performance metrics for the configuration in Table \ref{tbl:configuration-with-selected-hyper-parameter}}
    \begin{tabular}{|| c | c ||}
    \hline
         Metric & Score \\
         \hline
Training Loss & 0.6928 \\
Training Accuracy & 0.5117 \\
Validation Loss & 0.6929 \\
Validation Accuracy & 0.5117 \\
\hline
    \end{tabular}
    \label{tbl:performance-metrics-of-configuration-with-selected-hyper-parameter}
\end{table}
\raggedbottom

\begin{table}[htb]
\caption{Performance with different combinations of AE and DNN \cite{SewakARES}}
\centering
\small
\begin{tabular}{||c||c|c||c|c|c||c|c|c|}
\hline
\hline
DNN&AE&Accuracy&AE&Accuracy\\
\hline
\hline
2L-DNN & 1L-AE & 96.95 & 3L-AE & 97.16\\
4L-DNN & 1L-AE & 96.25 & 3L-AE & 99.21\\
7L-DNN & 1L-AE & 98.99 & 3L-AE & 93.60\\
\hline
\hline
\end{tabular}
 
\label{tbl:ae-dnn-results-120epochs}
\end{table}
\raggedbottom

\section{Conclusion} \label{sec:conclusion}
We conducted $\approx150$ experiments to investigate the relative importance (main effect) and the interaction-effect of different levels of various hyper-parameter covariates of an LSTM network viz. sequence/ padding length, embedding size, No. of LSTM layers, size of an LSTM layer, dropout ratio, batch-size, and optimizer algorithm for developing an effective AI based IDS. For this purpose, we used the opcode sequence features from the benchmark Malicia malware dataset. Careful analysis of the results from these experiments also indicates that there are significant interaction-effects amongst the different levels of various parameter covariates and the standard linear mechanism for hyper-parameter level selection is not optimal. Therefore, we also proposed an improved way by empirically controlling the overall importance of hyper-parameters with hierarchical selection ordered based on their main effect. 
From the results we found that the batch-size is one of the most important hyper-parameters that effects the selection of other parameters. Using our proposed methodology, we also concluded that the second and the third most influential parameter are the dropout ratio, and the padding length, followed by embedding length. These results are significant, because for LSTM networks as used for language-models, the focus had mostly been on increasing the number of layers to generate richer insights from text. Our work also validates that the inherent differences between opcode sequence and (word sequence) language is far from trivial from a perspective of porting of deep learning models and the assessment of their hyper-parameter importance.
We also compared the LSTM based IDS results to that obtained from the training of AE and DNN configuration which use (non-recurrent) opcode frequency-vector (instead of recurrent opcode sequence-vector) from the same benchmark dataset \cite{Malicia} and found that the various configurations used in our experiments did not produce a superior result. Therefore, we also inferred that for this dataset a recurrent DL architecture like the LSTM is not superior to non-recurrent DL architecture like the ML-DNN. This is a very significant finding, as often LSTMs as used in language-models, mostly outperform any algorithm that is based on non-sequential (and hence non-recurrent) representation of language, e.g., bag-of-words approach. This could be due to multiple reasons; the most obvious of them seems to be the exponentially longer sequence of opcodes as compared to sequence of tokens/ words in a sentence/ document, and due to the complex representation of malicious patterns within opcodes as compared to the representation of underlying meaning/concepts in the sequence of language tokens.
However, since opcode sequence are more complex than a sequence of language tokens, more records for different combination of levels across all evaluated parameters may help to further stabilize the results. Therefore, we plan to explore it more with a larger malware dataset.

\bibliographystyle{unsrt}
\bibliography{main}

\begin{thebibliography}{10}

\bibitem{lstm-language-modeling}
Martin Sundermeyer, Ralf Schl{\"u}ter, and Hermann Ney.
\newblock Lstm neural networks for language modeling.
\newblock In {\em Annual Conference of the International Speech Communication
  Association}, 2012.

\bibitem{sahay2020evolution}
Sanjay~K Sahay, Ashu Sharma, and Hemant Rathore.
\newblock Evolution of malware and its detection techniques.
\newblock In {\em Information and Communication Technology for Sustainable
  Development}, pages 139--150. Springer, 2020.

\bibitem{lstm-50-configurations}
Nils Reimers and Iryna Gurevych.
\newblock Optimal hyperparameters for deep lstm-networks for sequence labeling
  tasks.
\newblock {\em arXiv preprint arXiv:1707.06799}, 2017.

\bibitem{lstm-parameter-search}
Klaus Greff, Rupesh~K Srivastava, Jan Koutn{\'\i}k, Bas~R Steunebrink, and
  J{\"u}rgen Schmidhuber.
\newblock Lstm: A search space odyssey.
\newblock {\em IEEE Transactions on Neural Networks and Learning Systems},
  28(10):2222--2232, 2016.

\bibitem{Malicia}
Malicia {P}roject.
\newblock \url{ http://malicia-project.com}, 2012.
\newblock Accessed: Jan 2021.

\bibitem{SewakTENCON}
Mohit Sewak, Sanjay~K. Sahay, and Hemant Rathore.
\newblock Assessment of the relative importance of different hyper-parameters
  of lstm for an ids.
\newblock In {\em 2020 IEEE REGION 10 CONFERENCE (TENCON)}, pages 414--419.
  IEEE, 2020.

\bibitem{Sewak-DL-Introduction-DRL}
Mohit Sewak.
\newblock Introduction to deep learning.
\newblock In {\em Deep Reinforcement Learning}, pages 75--88. Springer, 2019.

\bibitem{LSTM-Language}
Wim De~Mulder, Steven Bethard, and Marie-Francine Moens.
\newblock A survey on the application of recurrent neural networks to
  statistical language modeling.
\newblock {\em Computer Speech \& Language}, 30(1):61--98, 2015.

\bibitem{LSTM-layer-selection}
Ming Tan, Cicero~dos Santos, Bing Xiang, and Bowen Zhou.
\newblock Lstm-based deep learning models for non-factoid answer selection.
\newblock {\em arXiv preprint arXiv:1511.04108}, 2015.

\bibitem{lstm-android-opcodes}
Jinpei Yan, Yong Qi, and Qifan Rao.
\newblock Lstm-based hierarchical denoising network for android malware
  detection.
\newblock {\em Security and Communication Networks}, 2018, 2018.

\bibitem{rathore2020detection}
Hemant Rathore, Sanjay~K Sahay, Shivin Thukral, and Mohit Sewak.
\newblock Detection of malicious android applications: Classical machine
  learning vs. deep neural network integrated with clustering.
\newblock In {\em International Conference on Broadband Communications,
  Networks and Systems}, pages 109--128. Springer, 2020.

\bibitem{Santos-2013}
Igor Santos, Felix Brezo, Xabier Ugarte-Pedrero, and Pablo~G Bringas.
\newblock Opcode sequences as representation of executables for
  data-mining-based unknown malware detection.
\newblock {\em Information Sciences}, 231:64--82, 2013.

\bibitem{rathore2020identification}
Hemant Rathore, Sanjay~K Sahay, Ritvik Rajvanshi, and Mohit Sewak.
\newblock Identification of significant permissions for efficient android
  malware detection.
\newblock In {\em International Conference on Broadband Communications,
  Networks and Systems}, pages 33--52. Springer, 2020.

\bibitem{SewakARES}
Mohit Sewak, Sanjay~K Sahay, and Hemant Rathore.
\newblock An investigation of a deep learning based malware detection system.
\newblock In {\em International Conference on Availability, Reliability and
  Security (ARES)}, pages 1--5, 2018.

\bibitem{DeepSign-2015}
Omid~E David and Nathan~S Netanyahu.
\newblock Deepsign: Deep learning for automatic malware signature generation
  and classification.
\newblock In {\em International Joint Conference on Neural Networks (IJCNN)},
  pages 1--8. IEEE, 2015.

\bibitem{Pascanu-2015}
Razvan Pascanu, Jack~W Stokes, Hermineh Sanossian, Mady Marinescu, and Anil
  Thomas.
\newblock Malware classification with recurrent networks.
\newblock In {\em IEEE International Conference on Acoustics, Speech and Signal
  Processing (ICASSP)}, pages 1916--1920. IEEE, 2015.

\bibitem{Saxe-2015}
Joshua Saxe and Konstantin Berlin.
\newblock Deep neural network based malware detection using two dimensional
  binary program features.
\newblock In {\em 10th International Conference on Malicious and Unwanted
  Software (MALWARE)}, pages 11--20. IEEE, 2015.

\bibitem{opcode-lstm-2layer}
Renjie Lu.
\newblock Malware detection with lstm using opcode language.
\newblock {\em arXiv preprint arXiv:1906.04593}, 2019.

\bibitem{microsoft-lstm}
Ben Athiwaratkun and Jack~W Stokes.
\newblock Malware classification with lstm and gru language models and a
  character-level cnn.
\newblock In {\em IEEE International Conference on Acoustics, Speech and Signal
  Processing (ICASSP)}, pages 2482--2486. IEEE, 2017.

\bibitem{Kolosnjaji-2016}
Bojan Kolosnjaji, Apostolis Zarras, George Webster, and Claudia Eckert.
\newblock Deep learning for classification of malware system call sequences.
\newblock In {\em Australasian Joint Conference on Artificial Intelligence},
  pages 137--149. Springer, 2016.

\bibitem{LSTM-api-call}
Sumith Maniath, Aravind Ashok, Prabaharan Poornachandran, VG~Sujadevi,
  Prem~Sankar AU, and Srinath Jan.
\newblock Deep learning lstm based ransomware detection.
\newblock In {\em Recent Developments in Control, Automation \& Power
  Engineering (RDCAPE)}, pages 442--446. IEEE, 2017.

\bibitem{LSTM-system-call}
Xi~Xiao, Shaofeng Zhang, Francesco Mercaldo, Guangwu Hu, and Arun~Kumar
  Sangaiah.
\newblock Android malware detection based on system call sequences and lstm.
\newblock {\em Multimedia Tools and Applications}, 78(4):3979--3999, 2019.

\bibitem{LSTM-PE-header}
Edward Raff, Jared Sylvester, and Charles Nicholas.
\newblock Learning the pe header, malware detection with minimal domain
  knowledge.
\newblock In {\em 10th ACM Workshop on Artificial Intelligence and Security},
  pages 121--132, 2017.

\bibitem{sewak2020doom}
Mohit Sewak, Sanjay~K Sahay, and Hemant Rathore.
\newblock Doom: a novel adversarial-drl-based op-code level metamorphic malware
  obfuscator for the enhancement of ids.
\newblock In {\em ACM International Conference on Pervasive and Ubiquitous
  Computing (UbiComp)}, pages 131--134, 2020.

\bibitem{SewakDRLDO}
Mohit Sewak, Sanjay~K Sahay, and Hemant Rathore.
\newblock Drldo: A novel drl based de-obfuscation system for defence against
  metamorphic malware.
\newblock {\em Defence Science Journal}, 71(1), 2021.

\bibitem{rathore2020robust}
Hemant Rathore, Sanjay~K Sahay, Piyush Nikam, and Mohit Sewak.
\newblock Robust android malware detection system against adversarial attacks
  using q-learning.
\newblock {\em Information Systems Frontiers}, pages 1--16, 2020.

\bibitem{LSTM-iot}
Hamed HaddadPajouh, Ali Dehghantanha, Raouf Khayami, and Kim-Kwang~Raymond
  Choo.
\newblock A deep recurrent neural network based approach for internet of things
  malware threat hunting.
\newblock {\em Future Generation Computer Systems}, 85:88--96, 2018.

\bibitem{LSTM-obfuscation}
Alessandro Bacci, Alberto Bartoli, Fabio Martinelli, Eric Medvet, and Francesco
  Mercaldo.
\newblock Detection of obfuscation techniques in android applications.
\newblock In {\em 13th International Conference on Availability, Reliability
  and Security (ARES)}, pages 1--9, 2018.

\bibitem{lstm-pe-75000}
Ben Athiwaratkun and Jack~W Stokes.
\newblock Malware classification with lstm and gru language models and a
  character-level cnn.
\newblock In {\em IEEE International Conference on Acoustics, Speech and Signal
  Processing (ICASSP)}, pages 2482--2486. IEEE, 2017.

\bibitem{lstm-android-1000-epochs}
R~Vinayakumar, KP~Soman, Prabaharan Poornachandran, and S~Sachin~Kumar.
\newblock Detecting android malware using long short-term memory (lstm).
\newblock {\em Journal of Intelligent \& Fuzzy Systems}, 34(3):1277--1288,
  2018.

\bibitem{lstm-hadamard-product}
Chandler Davis.
\newblock The norm of the schur product operation.
\newblock {\em Numerische Mathematik}, 4(1):343--344, 1962.

\bibitem{lstm-peephole-activation}
Felix~A Gers and E~Schmidhuber.
\newblock Lstm recurrent networks learn simple context-free and
  context-sensitive languages.
\newblock {\em IEEE Transactions on Neural Networks}, 12(6):1333--1340, 2001.

\bibitem{sewak-dl-overview}
Mohit Sewak, Sanjay~K Sahay, and Hemant Rathore.
\newblock An overview of deep learning architecture of deep neural networks and
  autoencoders.
\newblock {\em Journal of Computational and Theoretical Nanoscience},
  17(1):182--188, 2020.

\bibitem{lstm-padding-length}
Mahidhar Dwarampudi and NV~Reddy.
\newblock Effects of padding on lstms and cnns.
\newblock {\em arXiv preprint arXiv:1903.07288}, 2019.

\bibitem{box-plot}
Robert McGill, John~W Tukey, and Wayne~A Larsen.
\newblock Variations of box plots.
\newblock {\em The American Statistician}, 32(1):12--16, 1978.

\end{thebibliography}

\end{document}